\documentclass[prb,aps,twocolumn,nofootinbib]{revtex4}
\usepackage{epsf}

\newcommand{\eq}[1]{Eq.~(\ref{#1})}

\def\be{\begin{equation}}
\def\ee{\end{equation}}
\def\ba{\begin{eqnarray}}
\def\ea{\end{eqnarray}}
\def\half{{1 \over 2}}
\def\llangle{\langle\langle}
\def\rrangle{\rangle\rangle}
\def\vt{\vartheta}
\begin{document}
\title{Quantum mechanics of superconducting nanowires}
\author{S. Khlebnikov}
\affiliation{Department of Physics, Purdue University, West Lafayette,
Indiana 47907, USA}
\begin{abstract}
In a short superconducting nanowire connected to bulk superconducting leads, 
quantum phase slips behave as a system of linearly (as opposed to logarithmically)
interacting charges. This system maps onto quantum mechanics of a particle in
a periodic potential. We show that, while the state with a high
density of phase slips is not a true insulator (a consequence of Josephson 
tunneling between the leads), for a range of parameters it behaves as such down
to unobservably small temperatures. We also show that quantum phase slips give 
rise to multiple branches (bands) in the energy-current relation and to
an interband (``exciton'') mode.
\end{abstract}

\maketitle

\section{Introduction}
There is currently much interest in
low-temperature properties of very thin 
superconducting wires. The key process, significance of which one aims to understand,
is a quantum phase slip (QPS),\cite{Giordano} 
a virtual depletion of the superconducting density
through which the system tunnels to a different value of the supercurrent.
The rate of this process
depends not only on the ``bare'' fugacity---the tunneling exponential associated
with the phase slip core---but also on the interaction between individual QPS.

The present study has been motivated by the observation\cite{conf} that, for a wire
connected to bulk superconducting leads,
as the length 
of the wire is reduced, the logarithmic interaction between QPS crosses over to a 
linear one.
Since we are talking about a tunneling process, the interaction takes place in the 
$(x,\tau)$ plane, where $x$ is the spatial coordinate, and $\tau$ is the Euclidean time.
Heuristically,
the change in the character of the interaction comes about because,
during tunneling, the leads 
inhibit production of quasiparticles (plasmons) with nonzero wavenumbers, so 
the ``lines of force'' in the $(x,\tau)$ plane align in the 
$\tau$ direction and form a ``string'' with a nonzero tension.

Now, it is well known that in the case of a logarithmic
interaction, the system undergoes a phase transition that can be interpreted as 
unbinding of QPS--anti-QPS pairs. There are, in fact, several varieties of this 
phenomenon: a phase transition\cite{Anderson&al} in the Kondo problem
and the equivalent transition\cite{Chakravarty,Bray&Moore} in the two-state
dissipative quantum mechanics (DQM), the BKT transition\cite{Berezinskii,KT,Kosterlitz}
in the XY model and, finally, the superconductor-insulator 
transition (SIT) in the DQM with a periodic potential.\cite{Schmid,Bulgadaev}
Long superconducting wires with significant
amounts of disorder fall into the latter universality class.\cite{diso}
Is there a similar transition in short wires, where  the interaction between QPS 
is linear? Experiments do bring out a difference between 
short\cite{Bezryadin&al,Bollinger&al2007} and 
long\cite{Altomare&al,Zgirski&al2008} wires and 
suggest that a sharp transition exists in the former case. 

Here, we present two theoretical results concerning this problem. First,
we point out that in a short wire connecting bulk superconductors there
is no absolute distinction between a superconductor and an insulator: even
an ``insulator'' can support a weak supercurrent due to Josephson tunneling.
Second, we show that there is, nevertheless, a range of parameters for which the 
wire acts as an insulator down to unobservably small temperatures.

There is a mean-field-type
approach\cite{Meidan&al} to short wires that models QPS as an effective resistive
environment. In that approach, the renormalization-group equation for 
the QPS fugacity is taken to be the same as in a
long wire, except that the plasmon impedance is replaced by an effective 
shunt resistance. The latter is determined self-consistently and includes
the QPS channel ($R_{\rm ps}$), quasiparticles ($R_{\rm qp}$), and the impedance of 
the electrodes ($R_{\rm elec}$). 
While this procedure may qualitatively reflect the correct physics, it is not clear 
if it can be justified from first principles. Moreover, for the superconducting state
in the limit when $R_{\rm elec}= 0$ and $R_{\rm qp} \gg  R_{\rm ps}$, it does not 
reproduce the
exponential dependence of $R_{\rm ps}$ on the temperature found\cite{conf} by direct 
calculation.

Here, we consider the problem of SIT in short wires (with bulk superconducting leads)
starting directly from the description in terms of a classical gas of particles
with linear interactions in one ($\tau$) dimension (Sect.~\ref{sect:gas}). 
We study this system in
two different regimes (Sects.~\ref{sect:high} and \ref{sect:low}), 
where two 
different approximations are possible. In Sect. \ref{sect:crit} we pause to
describe two potentially observable effects predicted by our theory: a breakdown 
voltage, which, under certain conditions, provides
a direct measure of the QPS fugacity, and a curious ``exciton'' mode.
Prospects for detection of the breakdown voltage in experiment are discussed in 
Sect.~\ref{sect:exp}. Sect.~\ref{sect:concl} is a conclusion.

\section{One-dimensional gas with linear interactions} \label{sect:gas}
Our starting point is the
partition sum of linearly interacting charges, which represent QPS and anti-QPS.
This can be conveniently written as a path integral 
over an equivalent electrostatic potential $\phi(\tau)$:
\ba
\lefteqn{ Z = \int {\cal D} \phi(\tau) 
e^{-\frac{1}{2 K} \int d\tau (\partial_\tau \phi)^2 } \times }  \nonumber \\
& & \!\!\!\!\!\!\! \sum_{N_\pm =0}^\infty 
\frac{(\half\alpha)^{N_+ + N_-}}{N_+! N_-!} \int
\prod_{l=1}^{N_+}  d\tau_l e^{i \phi(\tau_l)}
\prod_{m=1}^{N_-}  d\tau'_m e^{-i \phi(\tau'_m)} .
\label{Z}
\ea
Note that the integral over the constant ($\tau$-independent) 
component of $\phi$ enforces the condition of charge neutrality: only terms with 
$N_+ = N_-$ contribute to the sum. Most of the expressions below will be for zero
temperature (i.e., infinite extent of the $\tau$ dimension), but occasionally we 
will discuss consequences of the cutoff on $\tau$ imposed by finite temperature.

The parameters appearing in \eq{Z} are $\alpha$, the bare QPS fugacity, and $K$,
the strength of the linear interaction. The role of $K$ can be understood by noting
that, for a single QPS--anti-QPS pair,
the path integral over $\phi(\tau)$ would be
\be
\int {\cal D} \phi(\tau)  
e^{-\frac{1}{2 K} \int d\tau (\partial_{\tau} \phi)^2  + i \phi(\tau_1) - i\phi(\tau_2)} 
=  e^{-\half K |\tau_1- \tau_2|} \; .
\label{pair}
\ee
These two parameters encode a wealth of microscopic details of the phase-slip process.
For example, if QPS are rare, $\alpha$ is proportional to $\exp(-S_{\rm core})$, where
$S_{\rm core}$, the action of the QPS core, depends on the device specifics,
such as the superconducting gap, the wire's width and thickness, and 
the width and thickness variations.
When the latter are significant, QPS may favor a constriction, making
$\alpha$ independent of the wire's length $L$. On the other
hand, in a uniform wire, QPS occur preferentially at the center,\cite{conf}
but the preference is rather weak, and $\alpha$ scales 
linearly with $L$.

The value of $K$ is determined by the one-dimensional superconducting stiffness 
$K_s(x)$ (including possible variations of $K_s$ with the width and thickness): 
\be
\frac{1}{K} = \frac{1}{4\pi^2} \int \frac{dx}{K_s(x)} \; .
\label{K}
\ee
Here $x$ is the coordinate along the wire. For a uniform wire, this gives 
$K \propto K_s/L$, i.e., $K$ scales
inversely with the length. The linear interaction in \eq{pair} then matches 
that obtained in Ref.~\onlinecite{conf}. We recall that, from the microscopic point
of view, the linear interaction between QPS reflects the energy of the initial
and final tunneling states, in which fluctuations of the magnitude of the order 
parameter are small. This is why it is possible to find the strength of the 
interaction using the phase-only theory.\cite{conf}

In more detail, the relation of \eq{Z} to the phase-only theory is as follows. 
Consider the field $\phi(x,\tau)$ dual 
to the  phase $\theta(x,\tau)$ of 
the superconducting order parameter in the sense that 
$\partial_\tau \theta \propto \partial_x \phi$. Since $\partial_x \phi$ 
is proportional to the charge density, $\phi$ can be thought of as the density
of the electric dipole moment (in suitable units).
When the leads are bulk superconductors,
all $\tau$-dependent components of $\theta$ satisfy the Dirichlet boundary 
conditions (b.c.) at the ends of the wire.\cite{conf}
In view of the duality relation above, the Dirichlet b.c.
for $\theta$ imply the Neumann b.c. for $\phi$, at both ends. 

Then, if $\alpha$ is much smaller than the charging energy of the wire, and 
the temperature is low enough, only the spatially uniform  mode of $\phi(x,\tau)$ 
matters, so
instead of a field we have a single quantum-mechanical degree of freedom,
$\phi(\tau)$. Leaving aside overall constant factors, one can think of it as
being the total dipole moment of the wire or, alternatively, the total charge
transported through the wire since some fixed initial moment of time. It also
coincides with the equivalent potential that appears in \eq{Z}.

\eq{Z} is an effective low-energy theory, which encodes the high-energy 
details in the values of its parameters, and so is valid only up to some cutoff
frequency $\Lambda$. Since the theory does not include fermionic quasiparticles,
the cutoff is of the order of or smaller than the superconducting gap $\Delta$.
Sometimes we will need a way to explicitly include such a cutoff in our 
calculations. One possibility, which we use in Sect.~\ref{sect:low},
is to modify the kinetic term of $\phi$ so as to suppress modes above a cutoff
frequency.
Another possibility, suitable for Monte Carlo simulations,
is to discretize time. Cutoff dependence is discussed
in Sect.~\ref{sect:exp}. For now, we consider the case when both $K$ and $\alpha$ 
are much smaller than $\Lambda$, so the cutoff does not matter.

The sums in \eq{Z} can be evaluated by noting that the sum over $N_+$ at fixed 
$N = N_+ + N_-$ is Newton's binomial, and the remaining sum over $N$ is an 
exponential (of $\alpha \int d\tau \cos\phi$). Thus, \eq{Z} is equivalent to the
theory with the Euclidean action
\be
S = \int d\tau \left\{ \frac{1}{2 K}  (\partial_\tau \phi)^2  - \alpha  \cos\phi(\tau) 
\right\}
\; .
\label{S}
\ee
This is similar to the mapping between the Coulomb gas and the sine-Gordon model
in two dimensions.\cite{Chui&Lee,Jose&al,Wiegmann}

\eq{S} is quantum mechanics of a particle in a cosine potential. Solutions to the 
corresponding Schr\"{o}dinger equation (the Mathieu functions) are known, 
and occasionally
we will refer to their properties. For most of the time, however, we will consider 
various limits of \eq{S}, for which the
recourse to the exact solution is not necessary.

Our system of charges with linear interactions is reminiscent of electrostatics in 
one dimension---an often used metaphor of quark confinement (see, for example,
Ref.~\onlinecite{Coleman&al}). As known in that context, depending on the dynamics 
of the charges, such a system can be in either a plasma or a confinement phase.
The distinction lies in the large-distance behavior of
a correlation function of {\em fractional} external charges. We therefore consider
\be
C_q(\tau) = \llangle e^{iq\phi(\tau)} e^{-iq\phi(0)} \rrangle
\label{corr}
\ee
with arbitrary $q$. Double brackets denote averaging with $\exp(-S)$. In a plasma
state, the charges can screen any
external charge, integer or fractional. This corresponds to \eq{corr} approaching 
a constant value at large $\tau$, for any $q$. In the confinement phase,
{\em integer} external charges
can still be screened, by new charges 
nucleating from the vacuum, but fractional charges cannot. As a result, 
the correlator (\ref{corr}) with fractional $q$ goes to zero at large $\tau$.

In the condensed-matter context, a plasma state, where the equivalent charges 
(the QPS) are unbound, corresponds to an insulator, and the
confinement phase to a superconductor.
We now argue that, in the present case, there is no true insulator, i.e., \eq{corr} 
always goes to zero for fractional $q$.

\section{High phase-slip density} \label{sect:high}
We begin with the limit that is the best candidate for an insulator:
\be
 \alpha \gg K \; ,
\label{cond1}
\ee
when QPS are abundant. In this limit, the dominant paths are those where
$\phi(\tau)$ spends most of the time in 
the vicinity of a minimum of the
cosine potential, but occasionally makes a transition to a neighboring minimum.

A natural way to describe this physics is the semiclassical approximation. To take into
account small fluctuations of $\phi$ near $\phi=0$, we expand 
the cosine in \eq{S} in powers of $\phi$ and treat the $\phi^2$ term as a ``mass'', and 
the higher powers as an interaction. The Green function of $\phi$ in the free massive 
theory
is
\be
G(\tau) = \frac{K}{2\omega_e} e^{-\omega_e |\tau|} \; ,
\label{G}
\ee
where $\omega_e$ is the characteristic frequency:
\be
\omega_e^2 = K \alpha \; .
\label{ome2}
\ee
Using this Green function as a propagator for perturbation theory, one finds that 
the perturbative expansion 
is controlled by the small parameter $K/\alpha$, and the asymptotics of the 
correlator (\ref{corr}) at large $\tau$, in perturbation theory, remains close to 1.

In addition to these perturbative fluctuations, however, there are large fluctuations
(instantons) connecting different minima of the cosine. They are solitons 
and anti-solitons of the sine-Gordon model, for example
\[
\phi_{\rm inst}(\tau) = 4 \arctan e^{\omega_e \tau} \; .
\]
Their action is
\be
S_{\rm inst} = 8 (\alpha / K)^{1/2} \; .
\label{Sinst}
\ee
These instantons are in a sense dual to the QPS, so when QPS are numerous, 
instantons are rare. Under the condition (\ref{cond1}),
they form a dilute gas with average density
\be
\bar{n} \sim \omega_e \sqrt{S_{\rm inst}} e^{-S_{\rm inst}} \; .
\label{ave_den}
\ee
Each instanton contained 
in the interval $(0,\tau)$ contributes $\exp(2\pi i q)$ to the correlator
(\ref{corr}), and each anti-instanton  $\exp(-2\pi i q)$. Statistics of these is that
of an ideal gas, i.e., is given by the Poisson distribution $P({\cal N})$. Hence,
\ba
C_q(\tau) & \approx & \sum_{{\cal N}_+ {\cal N}_-} P({\cal N}_+) P({\cal N}_-) 
e^{2\pi i q({\cal N}_+ - {\cal N}_-)}  \nonumber \\
\lefteqn{
= e^{-2 \tau \bar{n} [1- \cos(2\pi q) ]} \; . }
\label{inst}
\ea
We see that, due to the instantons, fractional charges are confined, with the string 
tension proportional to the instanton density. Note, however, that 
in practice $\tau$ is cutoff by the inverse temperature, 
so the confinement of fractional 
charges (i.e., superconductivity) will not be observed until
the temperature gets as low as $T \sim {\bar n}$, which in the present case is 
exponentially small. 

The underlying physics becomes more transparent if one observes that
$\partial_t \phi / 2\pi$ is the electric 
current in units of $2e$ (here $t = -i\tau$ is the real time, and $e$ 
is the electron charge). 
Thus, each instanton transports $2e$ of charge through 
the wire. A delocalized state of $\phi$ (a Bloch wave of the theory (\ref{S})) 
corresponds to a steady current.  At small
$K/\alpha$, the instanton rate is strongly suppressed, and for many purposes 
the system behaves as an insulator, but even then
a weak steady current is possible by tunneling of charges through
the wire (the Josephson effect).

Note also that, in a uniform wire, both $K$ and $\alpha$ depend on the wire's
length $L$, $\alpha$ being directly and $K$ inversely proportional to it. 
Decreasing the length connects smoothly the  ``insulating'' state of a nanowire to
the superconducting state of a Josephson junction (JJ).

\section{Exciton and the breakdown voltage} \label{sect:crit}

The spectrum of the theory (\ref{S}) has a band structure.
In the path integral formalism, the quasimomentum of these bands appears
as a $\vt$ angle associated with 
the instanton number
\[
Q = \frac{1}{2\pi} \int d\tau \partial_\tau \phi \; .
\]
Thus, we generalize \eq{S} as follows:
\be
S \to S + i \vt Q \; .
\label{Stheta}
\ee
Each instanton now contributes an additional factor of $e^{\pm i\vt}$, and
in the dilute-gas approximation the dependence of the partition sum on $\vt$ is 
\be
\frac{Z(\vt)}{Z(0)} = \exp [ 2 \int d\tau \bar{n} (\cos\vt-1 ) ] \; .
\label{Ztheta}
\ee
The average current is obtained by differentiating $\ln Z(\vt)$ with respect to
$\vt$ and equals
\be
I = \frac{i}{2\pi} \llangle \partial_\tau \phi \rrangle =  2 \bar{n} \sin \vt \; ,
\label{cur}
\ee
where $I$ is in units of $2e$. The maximal (critical) current is $I_c = 2 \bar{n}$.
\eq{cur} is of the familiar Josephson form, so we conclude that the quasimomentum
(or $\vt$ angle) of the $\phi$-description is precisely the relative phase
$\Delta \theta$ of the order parameter at the two ends of the wire.

The interpretation of $\Delta \theta$ as quasimomentum is particularly apt given 
that in the absence of phase slips (i.e., for $\alpha = 0$) $\Delta \theta$,
multiplied by the superfluid density, is the total momentum of the superfluid 
in the wire. One can say that phase slips turn momentum into quasimomentum.
This is similar to how phase slips enforce periodicity with respect to the 
magnetic flux enclosed by a superconducting nanoring.\cite{Matveev&al} 

\begin{figure}
\epsfxsize=3in \epsfbox{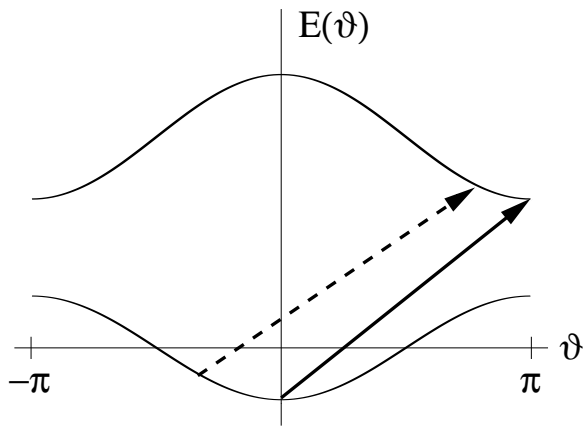}
\caption{A schematic of the band structure of the theory (\ref{S}). The arrows show
exciton transitions for two different values of the biasing current $I$. 
Note that the exciton frequency in general depends on $I$.}
\label{fig:bands}
\end{figure}

Note, though, that $\ln Z(\vt)$ gives only one band of the band structure, 
namely, the one corresponding to the ground state. In fact, there are additional bands,
which can be obtained from the exact solution to the quantum-mechanical problem
(\ref{S}). They
correspond to additional branches of energy as a function of the quasimomentum, 
$E(\vt)$, as shown in Fig.~\ref{fig:bands}.
One consequence of this is the existence of an interband excitation, 
which we term the exciton. 
At a fixed biasing current $I$, the exciton connects states for which the energy curves
have equal slopes $dE/d\vt = I$. Since these states have different values of $\vt$
(in particular, for $I=0$, the excited state is
a $\pi$-state), the transition produces a pulse of voltage.

The exciton is distinct from the usual Josephson
plasmon, and the band structure described above is distinct from that described
in Ref.~\onlinecite{Likharev&Zorin}. The plasmon can be included
in the theory by supplying $\vt$ 
with nonstatic modes and generalizing the action (\ref{Stheta}) into
\be
\widetilde{S} =  S + \frac{i }{2\pi} \int d\tau \vt \partial_\tau \phi
+  T \sum_{n\neq 0} P(\Omega_n) |\vt_n|^2 \; ,
\label{Stilde}
\ee
where $T$ is the temperature, $\Omega_n = 2\pi n T$ ($n$=integer) are the Matsubara
frequencies, and $P(\Omega)$ is the plasmon
response kernel. The interaction between the plasmon and the exciton, represented 
by the second term in \eq{Stilde} is the standard $\hat{I} \vt$ interaction, with
$\hat{I} \propto \partial_\tau \phi$ describing the fluctuating current.

The only nonlinearity in \eq{Stilde} is the cosine potential in $S$, \eq{S}.
If $\vt(\tau)$ were the slowest variable in the problem, we could have, in the
dilute instanton gas approximation, integrated $\phi$ out, just as we did for
a constant $\vt$ in \eq{Ztheta}. This is possible if, at small $\Omega$,
$P(\Omega)$ is purely capacitive, $P(\Omega) \propto C \Omega^2$, and 
the capacitance $C$ is sufficiently small. In this case, we recover the theory of 
Ref.~\onlinecite{Likharev&Zorin}. We know, however, that, in general,
superconducting leads
modify the low-frequency response significantly. For effectively one-dimensional 
leads, $P(\Omega) \propto |\Omega| / Z$ at small $|\Omega|$, $Z$ being the 
plasmon impedance of the leads (see Ref.~\onlinecite{conf}), while for 
three-dimensional ones we expect $P(\Omega)$ to go to a constant. The limit of
``bulk'' leads considered in this paper corresponds to $P$ being large, so that 
integrating out the $n\neq 0$ modes of $\vt$ 
results only in a small correction to the kinetic term in \eq{S}. This allows us
to neglect plasmons altogether in a study of the low-frequency (ground-state) 
properties. We note, however, that the plasmon-exciton coupling may become important 
when the device is operated at a higher frequency, e.g., by being 
subjected to microwave radiation.

Since $\phi$ 
is proportional to the dipole moment of the wire, it couples, via a $\phi V$
term, to the voltage $V$ across the wire. For a wire
deep in the ``insulating'' regime, $\phi$ can
sit for a long time near one minimum of the cosine potential. Its fluctuations there 
are small, and its value is related to the voltage by
\be
V = \frac{\pi \alpha}{e} \sin \phi \equiv V_c \sin \phi \; .
\label{V}
\ee
At small $V$, the exciton is a transition to an excited state near the same
minimum. The critical value $V_c = \pi \alpha  / e$ is the breakdown voltage, at
which $\phi$ begins to escape classically, and the system becomes a conductor.

Another breakdown channel, which opens at $V > V_g \equiv 2 \Delta/e$, is production 
of quasiparticle pairs. However, the condition $\alpha \ll \Lambda \alt \Delta$, 
under which we have been operating (and validity of which for 
realistic wires is discussed in Sect.~\ref{sect:exp}), guarantees that 
$V_c < V_g$.

\section{Low phase-slip density} \label{sect:low}
We now consider the limit of small QPS fugacity,
\be
\alpha \ll \min\{ K, \Lambda \} \; ,
\label{cond2}
\ee
where $\Lambda$ is the ultraviolet (UV) cutoff.
In this case, it is natural to expand the partition sum in powers of $\alpha$.
This brings in correlators of the Gaussian theory,
\be
S_0 = \frac{1}{2 K} \int d\tau  (\partial_\tau \phi)^2 \; ,
\label{S_0}
\ee
which need to be regulated in the infrared. To this end, we modify \eq{S_0} as follows
\be
S_0 \to 
\half \int d\tau d\tau' \phi(\tau) M_0(\Lambda, \mu; \tau, \tau') \phi(\tau') \; ,
\label{M_0}
\ee
where $M_0$ is a differential operator whose inverse (the Green function) 
$G_0 \equiv  M_0^{-1}$ is
\be
G_0(\Lambda, \mu; \tau, \tau') =  
\frac{K}{2} \left( 
\frac{1}{\mu} e^{-\mu |\tau-\tau'|} - \frac{1}{\Lambda} e^{-\Lambda |\tau-\tau'|}
\right) .
\label{G_0}
\ee
This implements an infrared (IR) cutoff at frequencies of order $\mu$ 
($\mu \ll \Lambda$). Indeed, for $\mu |\tau - \tau'| \ll 1$, \eq{G_0} nearly coincides 
with the unregulated Green function, proportional to $|\tau - \tau'| + \mbox{const}$,
but at large distances it decays exponentially, instead of growing linearly.
In addition, \eq{G_0} provides an explicit UV cutoff at frequencies of order $\Lambda$.

The requisite correlators are
\ba
\langle e^{iq \phi} \rangle & = & e^{-\half q^2 G_0(0)} \; , \\
\hspace{-20pt} \langle e^{iq \phi(0)} e^{\mp i \phi(\tau)} \rangle & = & 
\exp [ -\half (q^2 + 1) G_0(0) \pm q G_0(\tau) ] \; ,
\ea
etc. Single angular brackets denote averages in the Gaussian theory, 
and we have used the shorthand notation 
$G_0(\tau) \equiv  G_0(\Lambda, \mu; \tau, 0)$.
We do not include a $\vt$ angle in this section.

For simplicity, in this section we calculate, instead of the correlator 
(\ref{corr}), directly its large $\tau$ limit, i.e., the expectation value of the
``disorder parameter''
\be
\llangle e^{iq \phi(0)} \rrangle = \langle e^{iq \phi(0)} \rangle +
\alpha \int d\tau \langle e^{iq \phi(0)} \cos \phi(\tau) \rangle_c + \ldots ,
\label{op}
\ee
where the subscript $c$ denotes the connected part. The expectation
value is even in $q$, and in what follows we assume $q > 0$.

The first term on the right-hand side of \eq{op} rapidly vanishes as the IR
cutoff is removed ($\mu \to 0$), but the second one generally does not. 
The relevant piece is
\be
\int d\tau \langle e^{iq \phi(0)} e^{-i\phi(\tau)} \rangle_c
\approx \frac{4}{qK} e^{-\frac{K}{4\mu} (q - 1)^2 } f(K/\Lambda, q) \; ,
\label{piece}
\ee
where $f$ is a dimensionless function independent of $\mu$; $f(0, q) = 1$.
The approximation sign indicates that we have retained only the leading term
in the $\mu \to 0$ limit. We see that for any $q\neq 1$ \eq{piece} vanishes
exponentially at $\mu\to 0$, but for $q=1$ it remains constant. 

The pattern repeats itself in higher orders in $\alpha$, except that now 
larger (integer) charges can also be screened. This is again the physics 
of confinement. Note that the main contribution to the integral in \eq{piece}
for $q=1$
comes from small $|\tau|$, which means that the screening charge, represented
by $e^{-i\phi(\tau)}$ in \eq{piece}, nucleates quite close
to the external charge. For example, for $K \ll \Lambda$, 
the main contribution
comes from $|\tau| \sim 1/K$. This is how it should be since, according to
\eq{pair}, the string tension is $\half K$.

\section{Breakdown voltage in experiment} \label{sect:exp}
At temperatures $T \ll \Delta$ and with enough disorder, the ratio $K/\Delta$
depends on only one parameter---the total normal-state resistance of 
the wire $R_N$:\cite{conf}
\be
\frac{K}{\Delta} = \frac{2\pi^2 R_q}{R_N} \; .
\label{ratio}
\ee
Here $R_q = \pi / 2e^2 = 6.45$ k$\Omega$ is the resistance quantum. 
So, for a wire with $R_N \sim R_q$, the ratio is $K/\Delta \sim 20$.
Estimates suggest that the right quantity for numerical comparison with the UV
cutoff $\Lambda \alt \Delta$ is not $K$ itself but $\frac{1}{4} K$ or maybe even 
$\frac{1}{8} K$. In either case, though, such a wire is quite far from 
the cutoff-independent regime $K,\alpha \ll \Lambda$. 

To locate the crossover between the ``insulating'' state and conventional
superconductivity in the presence of a cutoff, we have used Monte Carlo simulations, 
with time discretized in steps of
$\Delta \tau = \pi / \Lambda$ and the crossover defined (somewhat arbitrarily)
by the condition $\llangle e^{i\phi} \rrangle = 0.5$. Preliminary data indicate 
that, at large $K/\Lambda$, the crossover occurs at a relatively small and nearly
$K$-independent value $\alpha_c \approx 0.4 \Lambda$. (Recall that
$\alpha$ is the fugacity per the entire wire, and in a uniform wire is proportional 
to the length). Therefore, we do not exclude that the observed zero-bias peaks in 
differential resistances of ``insulating'' wires\cite{Bollinger&al,Tinkham} 
are a manifestation of the electrical
breakdown effect described by \eq{V}, and the excess (``offset'') voltage contained
in such a peak provides an estimate of the breakdown voltage $V_c$.

\section{Conclusion} \label{sect:concl}
In this paper, we have presented a theory of short superconducting wires that uses, 
as its basic variable, the wire's electric dipole moment. We have found this 
description to be well suited to cases when the more conventional phase variable
has been made ``heavy'' (i.e., its fluctuations have become inhibited) by
sufficiently large, ``bulk'' superconducting leads.
The prominence of the dipole moment (polarization) means that
we view a small superconductor as if it were almost an insulator. We have seen, 
however, that, even in a thinnest wire, the true insulating behavior is prevented
by the Josephson tunneling between the leads. 

The thinner the wire is, the weaker that tunneling is. For sufficiently thin wires,
it is described by instantons of the theory (\ref{S}), which are suppressed
exponentially, so
the insulating behavior persists down to exponentially low, unobservably small 
temperatures. The height of the potential traversed by the instantons, $\alpha$
in \eq{S}, has the interpretation of the bare fugacity of quantum phase slips in the
wire.

We have presented two potentially testable predictions of our theory. One effect, 
which, as discussed in Sect.~~\ref{sect:exp}, may already have been observed, 
is the existence
of a breakdown voltage in ``insulating'' wires. The other is the exciton,
corresponding to the interband transitions in the theory (\ref{S}). It can presumably
be detected as a feature in the absorption spectrum. We should note,
however, that \eq{S} can be literally applied to computation of the exciton spectrum 
only when both $\alpha$ and the exciton frequency are much smaller 
than the UV cutoff $\Lambda$.

{\em Note added in v4.} Recently, we learned about the work of Mooij and 
Nazarov.\cite{Mooij&Nazarov} They start from a somewhat
different premise than we do, but their results overlap with ours.

\begin{acknowledgments}
The author thanks A. Bezryadin for useful comments.
\end{acknowledgments}

\end{document}